\documentclass[aip, cha, reprint]{revtex4-1}

\usepackage{amsmath}
\usepackage{graphicx}
\usepackage{epstopdf}
\usepackage[caption=false]{subfig}

\begin{document}

\title{Precession-torque-driven domain-wall motion in out-of-plane materials}

\author{M.J.G. Peeters}
\email[E-mail: ]{m.j.g.peeters@tue.nl}
\affiliation{Department of Applied Physics, Eindhoven University of Technology, PO Box 513, 5600 MB Eindhoven, The Netherlands}

\author{F.C. Ummelen}
\affiliation{Department of Applied Physics, Eindhoven University of Technology, PO Box 513, 5600 MB Eindhoven, The Netherlands}

\author{M.L.M. Lalieu}
\affiliation{Department of Applied Physics, Eindhoven University of Technology, PO Box 513, 5600 MB Eindhoven, The Netherlands}

\author{J.-S. Kim}
\affiliation{DGIST Research Center for Emerging Materials, DGIST, Daegu 42988, South Korea}

\author{H.J.M. Swagten}
\affiliation{Department of Applied Physics, Eindhoven University of Technology, PO Box 513, 5600 MB Eindhoven, The Netherlands}

\author{B. Koopmans}
\affiliation{Department of Applied Physics, Eindhoven University of Technology, PO Box 513, 5600 MB Eindhoven, The Netherlands}

\date{\today}

\begin{abstract}
Domain-wall (DW) motion in magnetic nanostrips is intensively studied, in particular because of the possible applications in data storage. In this work, we will investigate a novel method of DW motion using magnetic field pulses, with the precession torque as the driving mechanism. We use a one dimensional (1D) model to show that it is possible to drive DWs in out-of-plane materials using the precession torque, and we identify the key parameters that influence this motion. Because the DW moves back to its initial position at the end of the field pulse, thereby severely complicating direct detection of the DW motion, depinning experiments are used to indirectly observe the effect of the precession torque. The 1D model is extended to include an energy landscape in order to predict the influence of the precession torque in the depinning experiments. Although preliminary experiments did not yet show an effect of the precession torque, our calculations indicate that depinning experiments can be used to demonstrate this novel method of DW motion in out-of-plane materials, which even allows for coherent motion of multiple domains when the Dzyaloshinskii-Moriya interaction is taken into account.
\end{abstract}

\pacs{}

\maketitle

\section{Introduction}
\label{sec:introduction}
Ever since the proposal of the racetrack memory \cite{Parkin2008} there has been much interest in the topic of domain-wall (DW) motion \cite{Haazen2013, Emori2013, Thiaville2012a}. The conventional approaches to drive DWs use spin-polarized currents \cite{Haazen2013, Emori2013} or magnetic fields \cite{Mougin2007a, Thiaville2012a}. As magnetic fields typically cannot provide coherent DW motion, thus resulting in loss of data in data storage devices, spin-polarized currents are generally used to drive DWs in magnetic racetracks \cite{Yang2015, Parkin2015}. A disadvantage, however, is that the large currents that are necessary can cause breakdown of devices due to Joule heating \cite{You2006b}.

In this work we will describe a novel method of DW motion, where magnetic fields are used in such a way that coherent DW motion is possible. The field is applied along a hard axis, i.e. perpendicular to the magnetization in the domains, with the resulting precession torque as the driving mechanism behind the DW motion. This has already been demonstrated for in-plane materials \cite{Kim2014}, and in this work we will explore the use of this method in out-of-plane (OOP) materials. This is especially interesting for the use in data storage devices, as OOP materials can provide much larger storage densities than IP materials \cite{Parkin2015}. As no current is sent through the nanowires when using this method, the chance of device breakdown is expected to be lower than for current-driven devices. Therefore, this method of DW motion can be interesting for applications where device lifetime is important.

We will first introduce this new method of DW motion and show that it can be used to drive DWs in OOP materials using a simple model. Then the key parameters that influence the DW motion will be explored, focussing on modelling the depinning experiments that can be used to verify this method of DW motion. The preliminary experiments we performed will be discussed as well, followed by a review of some challenges for this new method of DW motion, including how coherent DW motion can be achieved using the Dzyaloshinskii-Moriya interaction. 

\section{Theory}
\label{sec:theory}
To explain how the precession torque can drive DW motion we start with the Landau-Lifshitz-Gilbert (LLG) equation \cite{Landau1935, Gilbert2004}, the most general description of magnetization dynamics:
\begin{equation}
\frac{\mathrm{d}\textbf{M}}{\mathrm{d}t} = -\gamma\mu_0\textbf{M}\times\textbf{H}_\mathrm{eff} + \frac{\alpha}{M_\mathrm{s}}\left(\textbf{M}\times\frac{\mathrm{d}\textbf{M}}{\mathrm{d}t}\right).
\label{eqn:LLG}
\end{equation}
Here, $\textbf{M}$ is the local magnetization, $\gamma$ the gyromagnetic ratio, $\mu_0$ the vacuum permeability, $\textbf{H}_\mathrm{eff}$ the effective field the magnetization experiences, $\alpha$ the Gilbert damping factor and $M_\mathrm{s} = |\textbf{M}|$. The first term describes the precession of the magnetization around the effective field, while the second term describes the damping that causes the eventual alignment of $\textbf{M}$ with $\textbf{H}_\mathrm{eff}$.

\begin{figure}[!ht]
\includegraphics[width=0.39\textwidth]{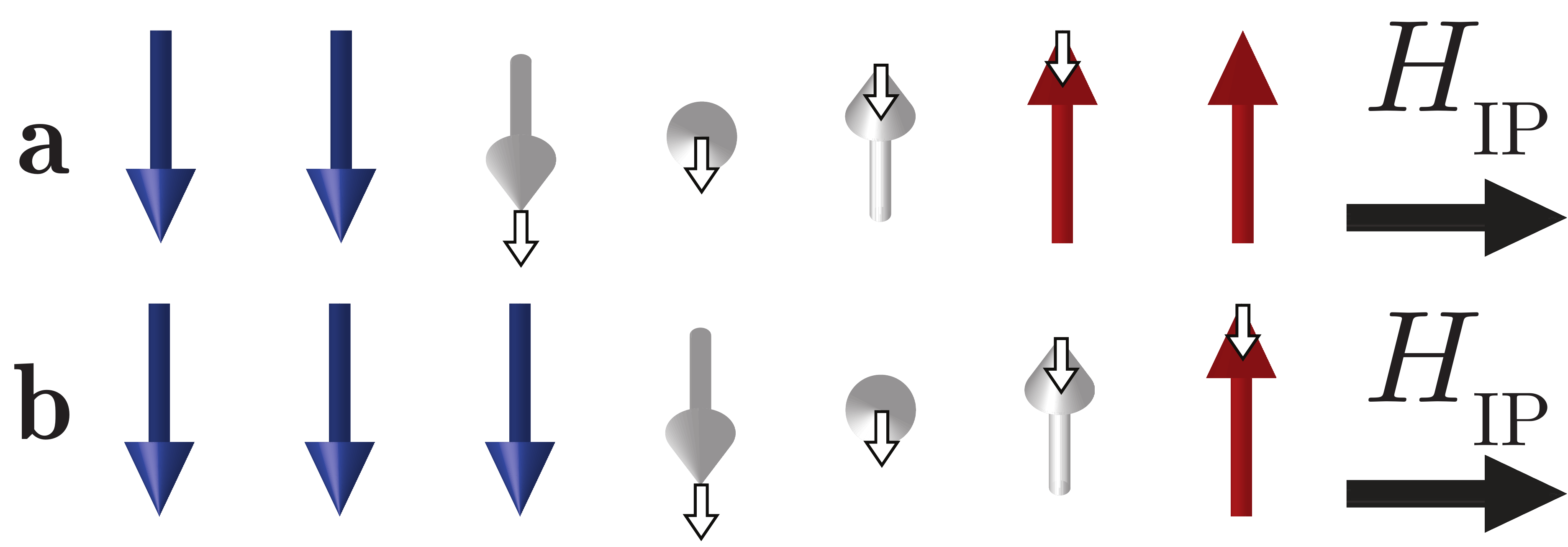}
\caption{\label{fig:explanation} \textbf{(a)} $H_\mathrm{IP}$ exerts a torque on the spins in the DW, causing the spins to rotate in the direction indicated by the small arrows. \textbf{(b)} The rotation of the spins causes an effective movement of the DW to the right.}%
\end{figure}

In Fig. \ref{fig:explanation} the working principle of precession-torque-driven DW motion is depicted. We start with a down domain and an up domain, separated by a Bloch DW. When an in-plane field $H_\mathrm{IP}$ is applied perpendicular to both the spins in the domains and the spins in the DW, the spins will experience a torque according to the precession term in the LLG equation. In the domains, both the anisotropy and the exchange energy prefer the spins to point in the same direction, while in the DW the anisotropy and the exchange energy are in competition. As a result of this the precession torque has a larger effect on the spins in the DWs than on the spins in the domains, which results in a rotation of the spins in the DW in the direction depicted by the small arrows in Fig. \ref{fig:explanation}a. In Fig. \ref{fig:explanation}b it can be seen that this rotation of the spins effectively moves the DW to the right, which is the essence of precession-torque-driven DW motion.

\section{Results}
\begin{figure}[!ht]
	\includegraphics[width=0.39\textwidth]{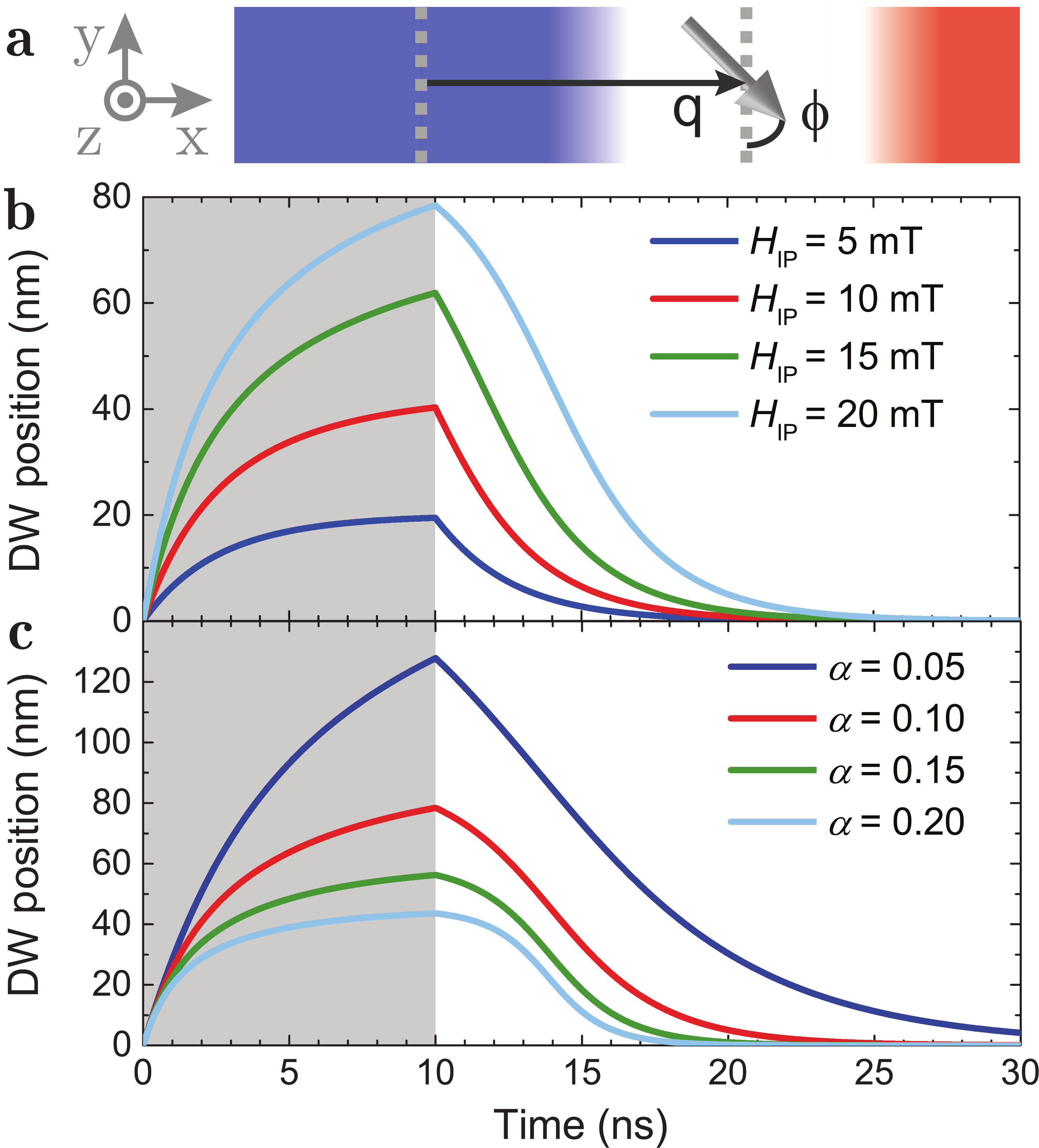}
    \caption{ \textbf{(a)} Top view of the DW, with $q$ the displacement of the DW and $\phi$ the internal DW angle. \textbf{(b)} DW displacement due to a 10 ns in-plane field pulse for various values of $H_\mathrm{IP}$ ($\alpha = 0.1$) and \textbf{(c)} for various values of $\alpha$ ($\mu_0 H_\mathrm{IP} = 20$ mT). }
    \label{fig:hxadependence}
\end{figure}

The DW motion can be described by deriving a one dimensional (1D) model, starting with the LLG equation \cite{Thiaville2004a}. In this model, the DW is defined by two collective coordinates, the DW position $q$ and the internal DW angle $\phi$, as defined in Fig. \ref{fig:hxadependence}a. By integrating the LLG equation over an infinitely long nanowire with a fixed DW profile \cite{Franken2012}, two equations are obtained that describe the two collective coordinates:
\begin{align}
\alpha \dot{q} - p \lambda \dot{\phi} &= \frac{\gamma\lambda}{2 M_\mathrm{s}}\left(\frac{\partial E}{\partial q}\right), \label{eqn:1dmodel1}\\
p \dot{q} + \alpha\lambda\dot{\phi} &= \frac{\gamma\lambda\mu_0}{2}\left[\vphantom{\frac{.}{.}}\pi \left(H_x \cos\phi + H_y\sin\phi\right) - H_\mathrm{D}\sin 2\phi\right]. \label{eqn:1dmodel2}
\end{align}
Here, $p$ determines the polarity of the DW, with $p = +1$ for an up-down DW and $p = -1$ for a down-up DW. $\lambda$ is the DW width, $H_\mathrm{D}$ is the anisotropy field of the DW, the sign of which determines whether Bloch or N\' eel walls are favored, and $\frac{\partial E}{\partial q}$ describes the energy landscape of the DW, including a possible magnetic field in the $z$-direction.

The parameters that are used in these simulations are realistic parameters for a Pt/Co/Pt stack \cite{Haazen2013}, with $\alpha = 0.1$, $K_\mathrm{eff} = 0.4875$ MJ/m$^3$, $\lambda = 5.7$ nm and $H_\mathrm{D} = 23$ mT. Using equations \ref{eqn:1dmodel1} and \ref{eqn:1dmodel2} the displacement of a DW due to a field pulse can be numerically calculated, as shown in Fig. \ref{fig:hxadependence}b and \ref{fig:hxadependence}c. In these figures the position of the DW during (grey area) and after (white area) a 10 ns field pulse is calculated for various field pulse strengths (Fig. \ref{fig:hxadependence}b) and various values of $\alpha$ (Fig. \ref{fig:hxadependence}c). It can be seen that after several nanoseconds the DW velocity decreases, which is due to the damping term in the LLG equation. This term causes the spins in the DW to eventually align with $H_\mathrm{IP}$, which reduces the torque ($\tau \propto \textbf{M}\times\textbf{H}_\mathrm{IP}$) and therefore impedes the DW motion, eventually causing the DW motion to stop. This is the reason there is a larger displacement for larger $H_\mathrm{IP}$ (larger torque) and for lower $\alpha$ (it takes longer for the spins to align with the applied field, which results in a large torque for a longer time). Furthermore, when $H_\mathrm{IP}$ is turned off there is an effective field in the $y$-direction, dominated by the demagnetization field of the DW. This creates a torque in the opposite direction of the torque generated by $H_\mathrm{IP}$, which causes the DW to move back to its initial position. In the outlook we will discuss how this backward motion can be eliminated, e.g. using pinning sites. 

\begin{figure}[!ht]
    \includegraphics[width=0.39\textwidth]{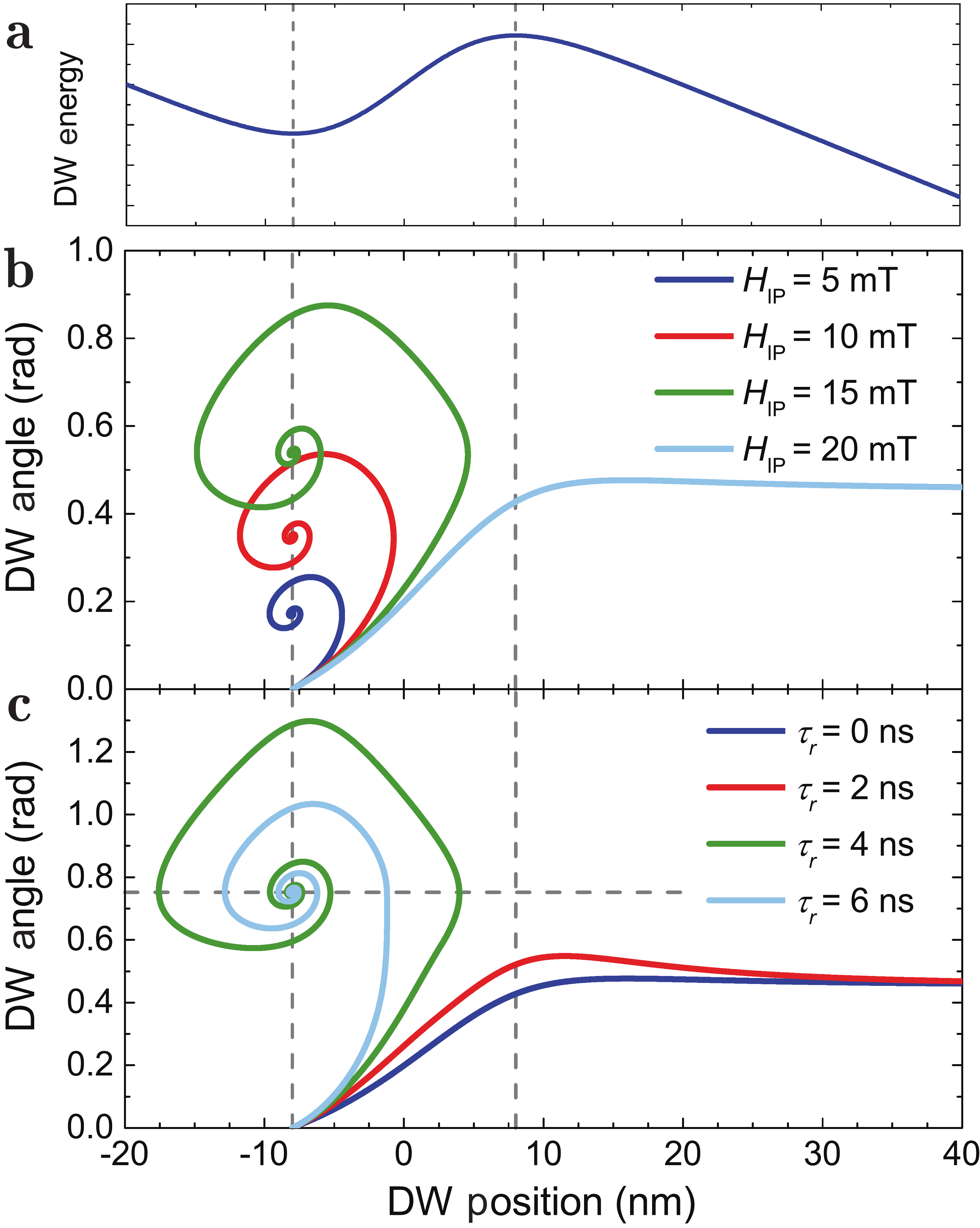}
    \caption{\textbf{(a)} DW energy landscape due to Ga$^+$-irradiation with $H_z < H_\mathrm{depin}$. \textbf{(b)} ($q$,$\phi$) diagrams for various values of $H_\mathrm{IP}$. The dashed vertical lines indicate the local minimum (left) and local maximum (right) in the energy landscape ($\tau_\mathrm{r} = 0$ ns). \textbf{(c)} ($q$,$\phi$) diagrams for various values of the rise time $\tau_\mathrm{r}$ ($H_x = 20$ mT).}
    \label{fig:phaseenergy}
\end{figure}

For the parameters we used, the DW displacement after a single field pulse is in the order of 100 nm. Combined with the fact that the DW moves back to its initial position after the pulse ends, this makes it difficult to detect the effect of the precession torque directly. With depinning experiments it is possible to overcome these complications, and detect the effect of the precession torque indirectly. For these experiments, Ga$^+$-irradiation can be used to create an energy barrier for the DW \cite{Franken2012}. This energy barrier can be overcome using a $z$-field, with the depinning field $H_\mathrm{dep}$ representing the critical $z$-field for which depinning happens. The precession torque can assist the depinning, and thus lower the depinning field. The energy landscape is incorporated in the 1D model via the $\frac{\partial E}{\partial q}$ term in equation \ref{eqn:1dmodel1}, and also includes a $z$-field, which causes the energy landscape to tilt. In Fig. \ref{fig:phaseenergy}a the energy landscape is shown for $H_z < H_\mathrm{dep}$. Assuming a constant effective anistropy in the irradiated (non-irradiated) region of $K_\mathrm{eff}$ ($K_\mathrm{eff,0}$), and a linear tranisition region with width $\delta$, the derivative of the energy of the system with respect to the DW position is given by \cite{Franken2012}
\begin{equation}
\frac{\partial E}{\partial q} = \frac{2\lambda}{\delta} \frac{(K_\mathrm{eff,0} - K_\mathrm{eff})\sinh\left(\frac{\delta}{\lambda}\right)}{\cosh\left(\frac{2q}{\lambda}\right) + \cosh\left(\frac{\delta}{\lambda}\right)} - 2\mu_0 M_\mathrm{s}H_z.
\end{equation}
In Fig. \ref{fig:phaseenergy}b and \ref{fig:phaseenergy}c it can be seen how an in-plane field pulse can cause the DW to depin for a $z$-field lower than the depinning field. The information is presented in a phase diagram, where the two degrees of freedom ($q$, $\phi$) are plotted against each other, in Fig. \ref{fig:phaseenergy}b for various values of $H_\mathrm{IP}$, in Fig. \ref{fig:phaseenergy}c for various values of the rise time of the IP field pulse. In both cases, initially the DW is positioned in the local energy minimum. The in-plane field then causes the DW to move to the right (in the direction of the energy barrier), and causes $\phi$ to increase towards the equilibrium angle $\phi_\mathrm{eq}$ (indicated by the horizontal dashed line in Fig. \ref{fig:phaseenergy}c) determined by $H_\mathrm{IP}$ and $H_\mathrm{D}$. When the in-plane field is strong enough (and the rise time is short enough), the DW will overcome the barrier before $\phi$ reaches $\phi_\mathrm{eq}$, and the DW depins. When $\phi$ reaches $\phi_\mathrm{eq}$ before that, the DW will move back due to the energy landscape. Eventually the DW will reach the equilibrium position ($q_0$, $\phi_\mathrm{eq}$) via the spiralling motion visible in Fig. \ref{fig:phaseenergy}b and \ref{fig:phaseenergy}c. The dependence of depinning on the rise time of the IP field pulse, as seen in Fig. \ref{fig:phaseenergy}c, is related to the effective field, determined by $H_\mathrm{D}$ and $H_\mathrm{IP}$. A long rise time results in an effective field that only slowly moves away from the direction of $H_\mathrm{D}$, the initial direction of the DW magnetization. This way, the magnetization can follow the effective field, and the torque is reduced ($\tau \propto \textbf{M}\times\textbf{H}_\mathrm{eff}$). Thus, to maximize the torque the field should reach its maximum value as quick as possible, corresponding with a short rise time. For long rise times the total DW displacement decreases, similar to low in-plane fields and high damping parameters. 

\begin{figure}[!ht]
    \includegraphics[width=0.39\textwidth]{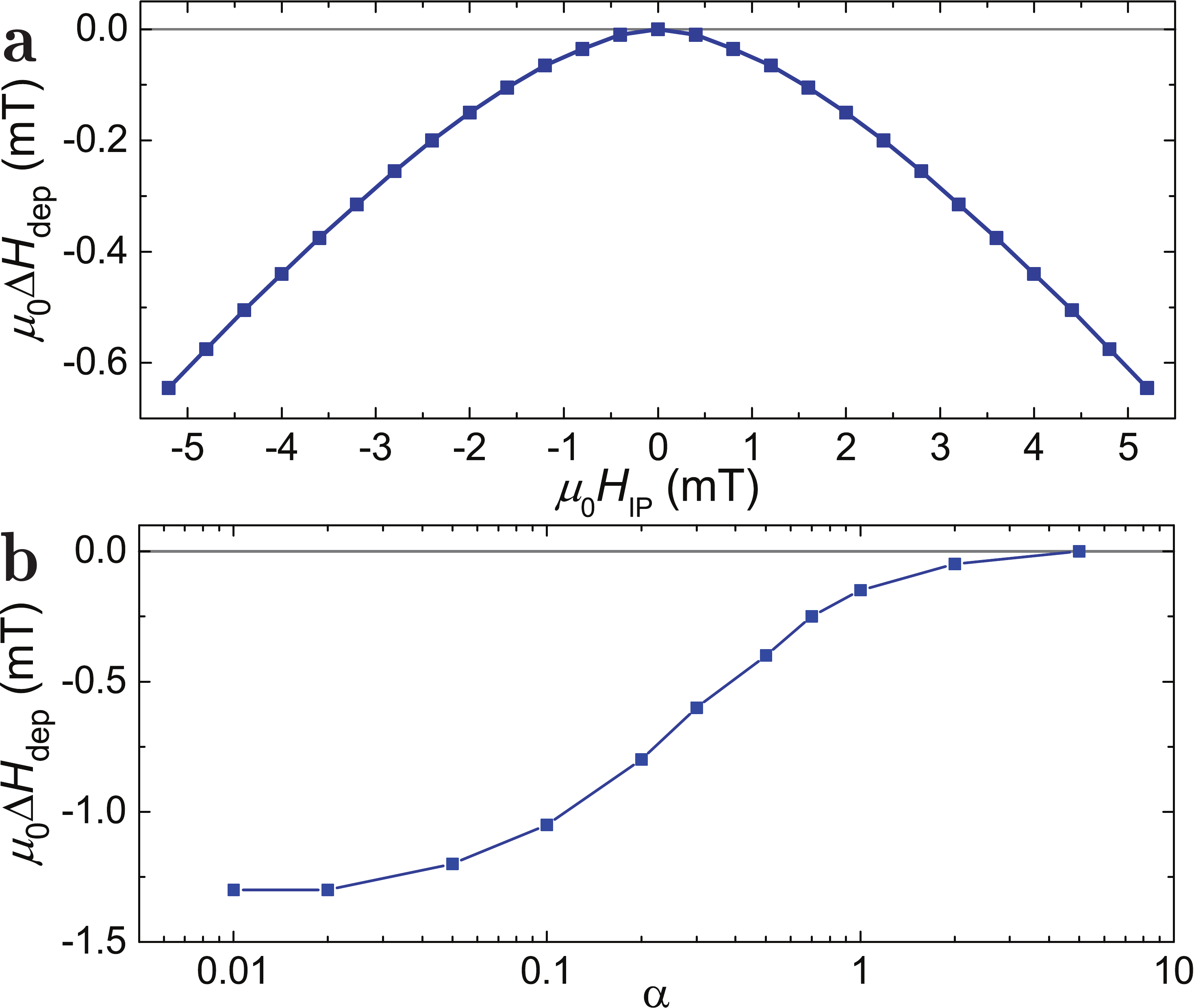}
    \caption{\textbf{(a)} Calculated change in depinning field due to in-plane field pulses. \textbf{(b)} Dependence of the change in depinning field at $H_\mathrm{IP} = 5$ mT on $\alpha$.}
    \label{fig:depinningdhdep}
\end{figure}

Experimentally, it is the depinning field that can be measured, e.g. with a Kerr microscope. Therefore, the 1D model is used to calculate how much the depinning field will change for various IP field strengths, the result of which is shown in Fig. \ref{fig:depinningdhdep}a. As expected, a stronger in-plane field corresponds to a larger change in the depinning field. More surprisingly, for both positive and negative in-plane fields there is a reduction of the depinning field. This is due to the fact that the torque generated by the in-plane field is in opposite directions at the start of the pulse and at the end of the pulse. Therefore, either at the start of the pulse or at the end of the pulse the torque is in the right direction to assist the depinning of the DW. In Fig. \ref{fig:depinningdhdep}b the dependence of the change in depinning field on the damping parameter is shown, for $H_\mathrm{IP} = 5$ mT. It is clear that the change in depinning field decreases for larger $\alpha$, in correspondence with the result from Fig. \ref{fig:hxadependence}b. 

We have performed preliminary experiments where we tried to show the effect of the precession torque on DW motion. The IP fields in the experiments were generated by sending a 10 ns current pulse through a gold waveguide. On top of this waveguide an insulating SiO$_2$ layer was deposited, followed by Ta(5 nm)/Pt(4 nm)/Co(0.6 nm)/Pt(4 nm) nanostrips ($1\times 10$ $\mu$m). The middle part of the nanostrips was irradiated with Ga$^+$-ions to introduce the DW energy barriers, and a Kerr microscope was used to determine the depinning fields. Although our calculations show that a change in depinning field of about 0.6 mT is expected for $\mu_0 H_\mathrm{IP} = 5$ mT, in these preliminary experiments no change could be detected. A possible explanation could be the value of $\alpha$, as we saw earlier that $\alpha$ has a large influence on the DW motion. TR-MOKE measurements of $\alpha$ \cite{Schellekens2013} show a field dependent value of $\alpha$, with $\alpha \approx 1$ for the fields used in the experiments and a decreasing $\alpha$ for higher fields. This field dependence can be explained by extrinsic contributions to the damping that disappear for high fields, such as inhomogeneous broadening\cite{Schellekens2013}. As visible in Fig. \ref{fig:depinningdhdep}b these high values of $\alpha$ make it more difficult to detect the effect of the precession torque in depinning measurements.

\section{Outlook}
\label{sec:outlook}
\begin{figure}[!ht]
\includegraphics[width=0.48\textwidth]{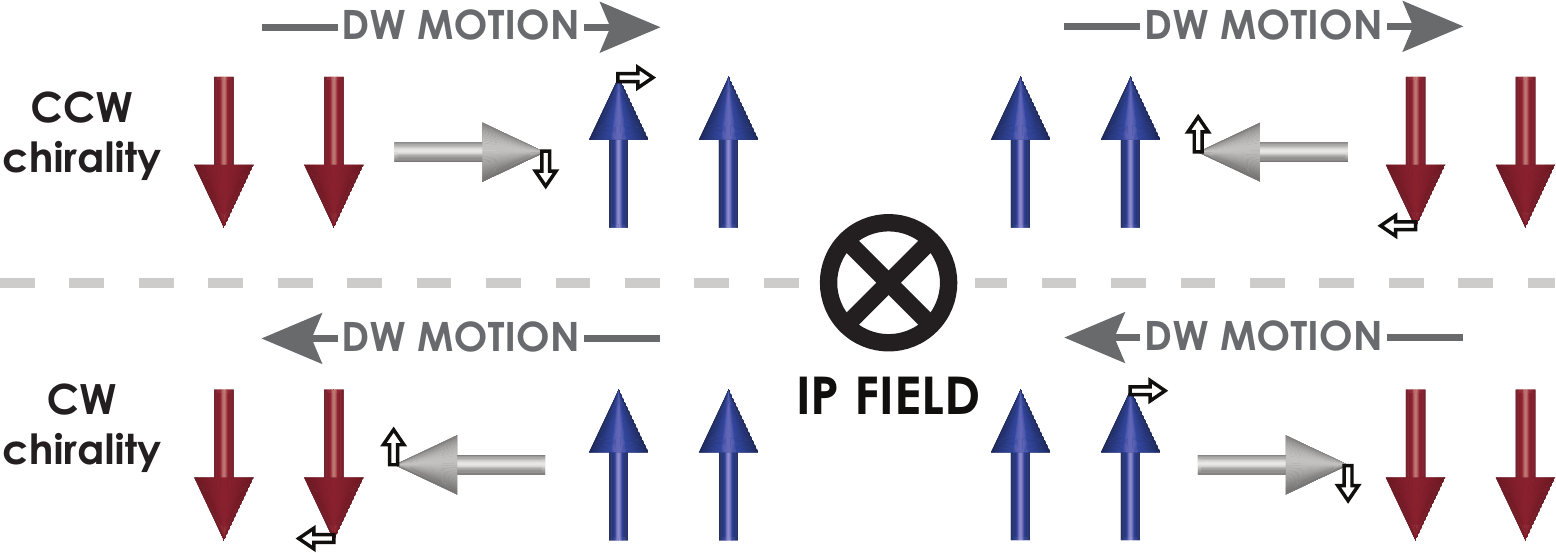}%
\caption{\label{fig:chiralitypolarity} Direction of DW motion for clockwise (CW) and counterclockwise (CCW) chiralities, for both polarities.}%
\end{figure}

Although we have shown theoretically that the precession torque can be used to drive DWs in OOP materials, there are still several issues that complicate the potential use in data storage devices. First, in a potential racetrack memory the reversal of the DW motion after the end of an in-plane field pulse inhibits any effective DW motion. A way to overcome this is the use of pinning sites \cite{Kim2014}. As the depinning from a pinning site depends on the rise or fall time of the field pulse, it is possible to adjust the rise and fall time in such a way that the DW depins during the rise time, but stays pinned at the next pinning site during the fall time, thus preventing the backward motion of the DW. A second issue we have not yet discussed is coherent motion of the DWs. To ensure coherent DW motion, a fixed chirality of the DWs is essential. The chirality defines whether the magnetization rotates in a clockwise or counterclockwise direction when passing through the DW in the positive $x$-direction. The Dzyaloshinskii-Moriya interaction, an anti-symmetric exchange interaction that prefers neighboring spins to be at an angle \cite{Dzyaloshinskii1958, Moriya1960}, favors N\' eel walls with a fixed chirality, which means that using materials with high DMI can ensure coherent DW motion with the precession torque. In Fig. \ref{fig:chiralitypolarity} the direction of DW motion for both down-up (left) and up-down (right) DWs can be seen for both chiralities, with the IP field now along the $y$-axis, perpendicular to the spins in the DWs. Indeed, as long as the chirality is fixed the DWs will move in the same direction, regardless of polarity, which ensures the required coherent DW motion. 

To conclude, we have used a 1D model to show that the precession torque can be used to drive DWs in OOP materials. Because of the backward motion of the DW at the end of the pulse, it is challenging to directly detect the DW motion experimentally. Therefore, we focussed on depinning experiments, for which the 1D model was extended with an appropriate energy landscape. This enabled us to predict the change in depinning fields as a result of the in-plane field pulses, with a dependence on $H_\mathrm{IP}$, $\alpha$ and the rise and fall time of the field pulse. Although we have not been able to measure the effect of the precession torque experimentally, possibly due to a high $\alpha$, our calculations indicate that it is feasible to use depinning experiments to observe the effect of the precession torque in OOP materials. 


\begin{acknowledgments}
The work is part of the research programme of the Foundation for Fundamental Research on Matter (FOM), which is part of the Netherlands Organisation for Scientific Research
(NWO), and the Gravitation program 'Research Centre for Integrated Nanophotonics', which is financed by the NWO.
\end{acknowledgments}

\bibliography{Papers-2016_MMM}

\end{document}